\documentclass[sensors,article,accept,moreauthors,pdftex]{mdpi} 

\firstpage{1} 
\pubvolume{21}
\issuenum{16}
\articlenumber{5590}
\pubyear{2021}
\copyrightyear{2021}
\datereceived{26 July 2021} 
\dateaccepted{15 August 2021} 
\datepublished{19 August 2021} 
\hreflink{https://doi.org/10.3390/s21165590} 
\pdfoutput=1

\Title{SG-WAS: a new Wireless Autonomous Night Sky Brightness Sensor}

\TitleCitation{SG-WAS: A New Wireless Autonomous Night Sky Brightness Sensor}

\Author{Miguel R. Alarcon $^{1, 2, *}$ \orcidA{}, Marta Puig-Subirà $^{1, 2, 3}$\orcidD, Miquel Serra-Ricart $^{1, 2}$\orcidB{}, Samuel Lemes-Perera $^{1}$\orcidC{}, Manuel Mallorquín $^{1,2,3}$\orcidE{}
 and César López $^{3, 4}$}

\AuthorNames{Miguel R. Alarcon, Marta Puig-Subirà, Miquel Serra-Ricart, Samuel Lemes-Perera, Manuel Mallorquín and César López}

\AuthorCitation{Alarcon, M.R.; Puig-Subirà, M.; Serra-Ricart, M.; Lemes-Perera, S.; Mallorquín, M.; López, C.}

\address{$^{1}$ \quad Instituto de Astrofísica de Canarias, C/ Vía Láctea s/n, E-38205 La Laguna, Canarias, Spain.\\

$^{2}$ \quad Departamento de Astrofísica, Universidad de La Laguna (ULL), E-38206 La Laguna, Canarias, Spain.\\

$^{3}$ \quad Sieltec Canarias S.L., C/ Hábitat No. 2, Portal D, Of. 3, E-38204 La Laguna, Canarias, Spain.\\

$^{4}$ \quad Departamento de Ingeniería Agraria, Naútica, Civil y Marítima, Universidad de La Laguna (ULL), E-38200 La Laguna, Canarias, Spain.}

\corres{Correspondence: mra@iac.es}

\abstract{The main features of SG-WAS (SkyGlow Wireless Autonomous Sensor), a low-cost device for 
measuring Night Sky Brightness (NSB), are presented. SG-WAS is based on the TSL237 sensor --like the Unihedron Sky Quality Meter (SQM) or the STARS4ALL Telescope Encoder and Sky Sensor (TESS)--, with wireless communication (LoRa, WiFi, or LTE-M) and solar-powered rechargeable batteries. Field tests have been performed on its autonomy, proving that it can go up to 
20 days without direct solar irradiance and remain hibernating after that for at least \mbox{4 months}, 
returning to operation once re-illuminated. A new approach to the acquisition of average NSB 
measurements and their instrumental uncertainty (of the order of thousandths of a magnitude) is presented. 
In addition, the results of a new Sky Integrating Sphere (SIS) method have shown the possibility 
of performing mass device calibration with uncertainties below 0.02 mag/arcsec$^2$. SG-WAS is the 
first fully autonomous and wireless low-cost NSB sensor to be used as an independent or networked 
device in remote locations without any additional infrastructure.}


\begin{document}
\section{Introduction}
{T}he increasing use of the artificial light at night (ALAN) has a 
dangerous impact on natural ecosystems. It can be so faint that  humans cannot see it, but~it has been shown that it could still threaten 30\% of 
vertebrates and 60\% of invertebrates that are nocturnal and very sensitive to light 
\citep{Holker2010, Gaston2013, Bennie2016, Owens2018}. {Eighty percent of the world's population lives in places where there is artificial light pollution, and about one-third of them cannot see the Milky Way. There are very few places left on the planet where natural darkness can be appreciated, observed, and measured} \citep{Falchi2016}. The~environmental effects of ALAN are not restricted to the direct nocturnal ecosystem impact; there is also 
some evidence of the role of outdoor lighting in the nocturnal dynamics of chemical pollutants over cities \citep{Stark2011}.

The emissions of ALAN from cities {can be} monitored using instruments on Earth-orbiting 
satellites \citep{SanchezdeMiguel2020, Jiang2018}. However, evaluating the effects of 
these emissions on night sky brightness (NSB) in {very} dark places (normally natural protected 
areas or astronomical observatories) requires extensive ground-based observations carried out 
by photometer networks \citep{Bara2018, Alarcon2021}. These networks are also necessary for testing light pollution models that predict NSB as a function of the location and distribution of artificial sources and the scattering of light in the atmosphere \citep{Aube2020}. 

\textls[-15]{The Unihedron Sky Quality Meter (hereafter SQM), a~low-cost and pocket-size NSB photometer, 
has given the general public the possibility of quantifying the NSB at any place and time 
\citep{Cinzano2007}. It is a silicon photodiode with a TSL237 light-to-frequency converter and a Hoya CM-500 filter that limits its effective bandpass to 400--650 nm. Based on the same TSL237 photodiode SQM detector, the~European funded project 
STARS4ALL\footnote{\url{www.stars4all.eu}} developed a new Telescope Encoder and Sky 
Sensor (\mbox{TESS~\cite{Zamorano2017}}), with~a more extended spectral response in the red (400--800 nm) 
to include the emission lines of High Pressure Sodium (HPS) lamps.}

Irradiance-to-frequency converter sensor technologies used in SQM and TESS have significantly 
expanded the range of inexpensive and easy-to-use NSB photometers, without~sacrificing the
 precision and detail of professional instruments \citep{Bar2019}. However,~although the cost 
of photometers has dropped, their deployment in remote locations (e.g.,\ protected natural 
areas) remains a major drawback for many projects. SG-WAS, a~new, low-cost NSB photometer 
based on the TSL237 sensor, is presented in this article, which is organized as follows. In~
Section~\ref{sec:methods}, a description of the SG-WAS photometer is given, together with a 
detailed explanation of WAS technology and field tests. The~new Sky Integrating 
Sphere (SIS) method and its results are presented in Section~\ref{sec:results}, along with a brief 
discussion of the implications of the results and uncertainties. Finally, the~main 
conclusions are provided in Section~\ref{sec:conclusion}.

\section{SG-WAS~Photometer} \label{sec:methods}
The SkyGlow Wireless Autonomous Sensor (SG-WAS) is a new NSB photometer partially funded by the EU 
project EELabs, and~designed and manufactured by the  Canarian R\&D company Sieltec Canarias S.L., 
under the scientific coordination of the Instituto de Astrofísica de Canarias (IAC, Tenerife, Spain). 
SG-WAS is basically composed~of the following.
\begin{enumerate}
    \item {A TSL237 irradiance-to-frequency converter sensor.}
    \item {An optical dichroic that determines the spectral response of the device.}
    \item {A concentrator optic that focuses the light and determines the effective field of view~(FOV).}
    \item {An infrared (IR) sensor that measures the sky temperature.}
    \item {\textls[-15]{Two microcontrollers that convert the frequencies measured into an average
magnitude and uncertainty, store and send the information, and~control the power-saving strategy.}}
    \item{\textls[-20]{A  communication unit  (LoRa, WiFi, LTE-M) customizable to the place where it is installed.}}
\end{enumerate}

\begin{figure}[t]
\includegraphics[width=13.5 cm]{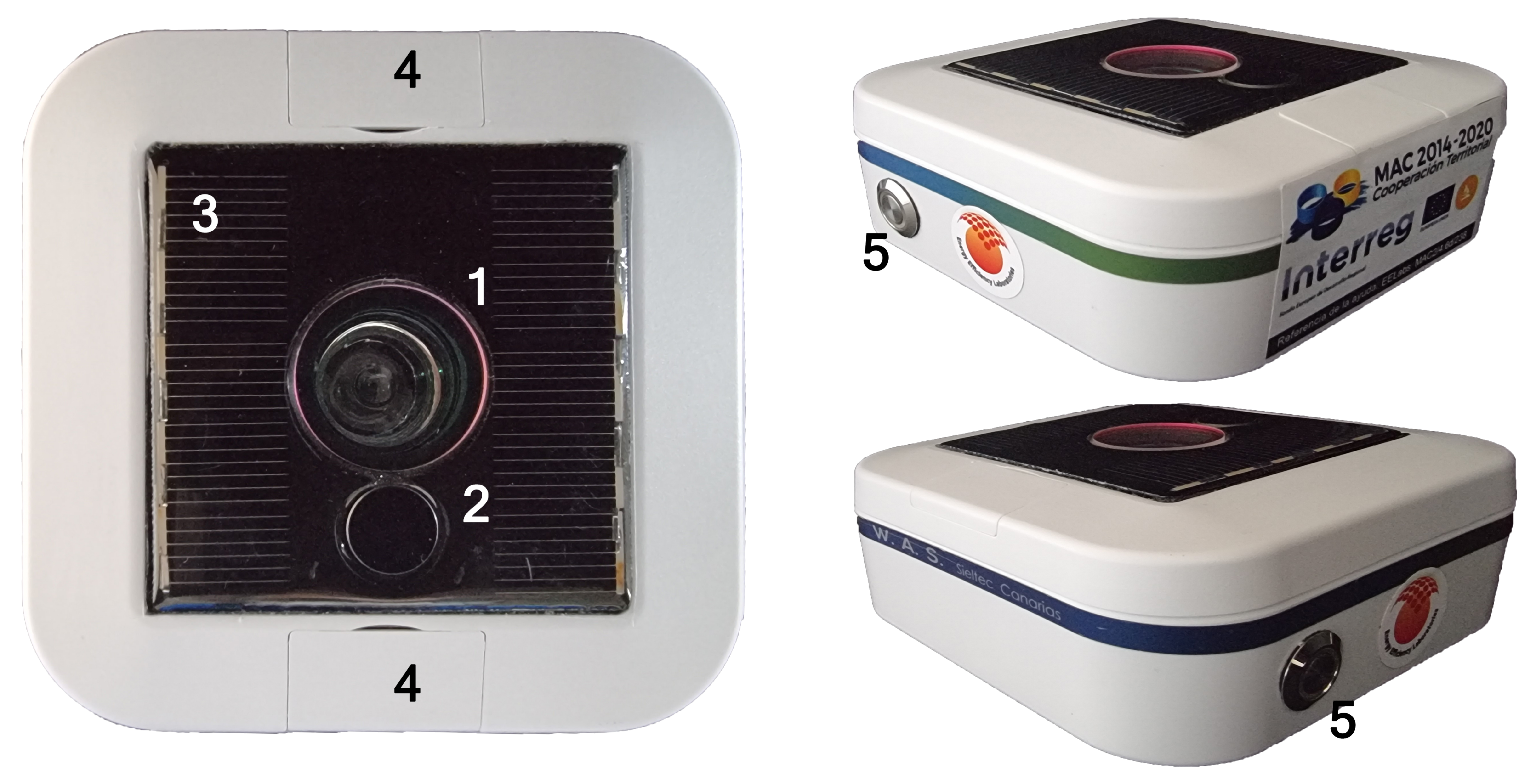}
\caption{{SG-WAS photometer viewed from above (\textbf{a}) and two sides (\textbf{b}). Its dimensions are 10 cm $\times$ 10 cm $\times$ 4 cm and it weighs about 100 g. The~main elements are   sky brightness optics (1), infrared thermometer (2), solar panel (3), fixing screws (4), and power button (5).}}
\label{fig:SG}
\end{figure}

The photometer's dimensions are similar to those of other photometers of its class \mbox{(10 cm $\times$ 10 cm $\times$ 4 cm)} and its
 weight is approximately 100 g. It is designed to be placed horizontally on a flat surface (optionally 
provided by the manufacturer), so its 20 mm-diameter window and the IR sky temperature 
sensor face the zenith {(see Figure~\ref{fig:SG})}. On~one side of its casing, there is a color-coded manual switch-off 
button that may be enabled in configuration mode. A~photovoltaic cell on the upper part 
allows it to be recharged within a few hours with direct sunlight (see Section~\ref{sec:Autonomous}). 
The SG-WAS is the first fully autonomous and wireless NSB photometer. It is waterproof and resistant 
to adverse weather~conditions.

\subsection{Wireless}
\label{sec:Wireless}
\begin{figure}[t]
\includegraphics[width=13.5 cm]{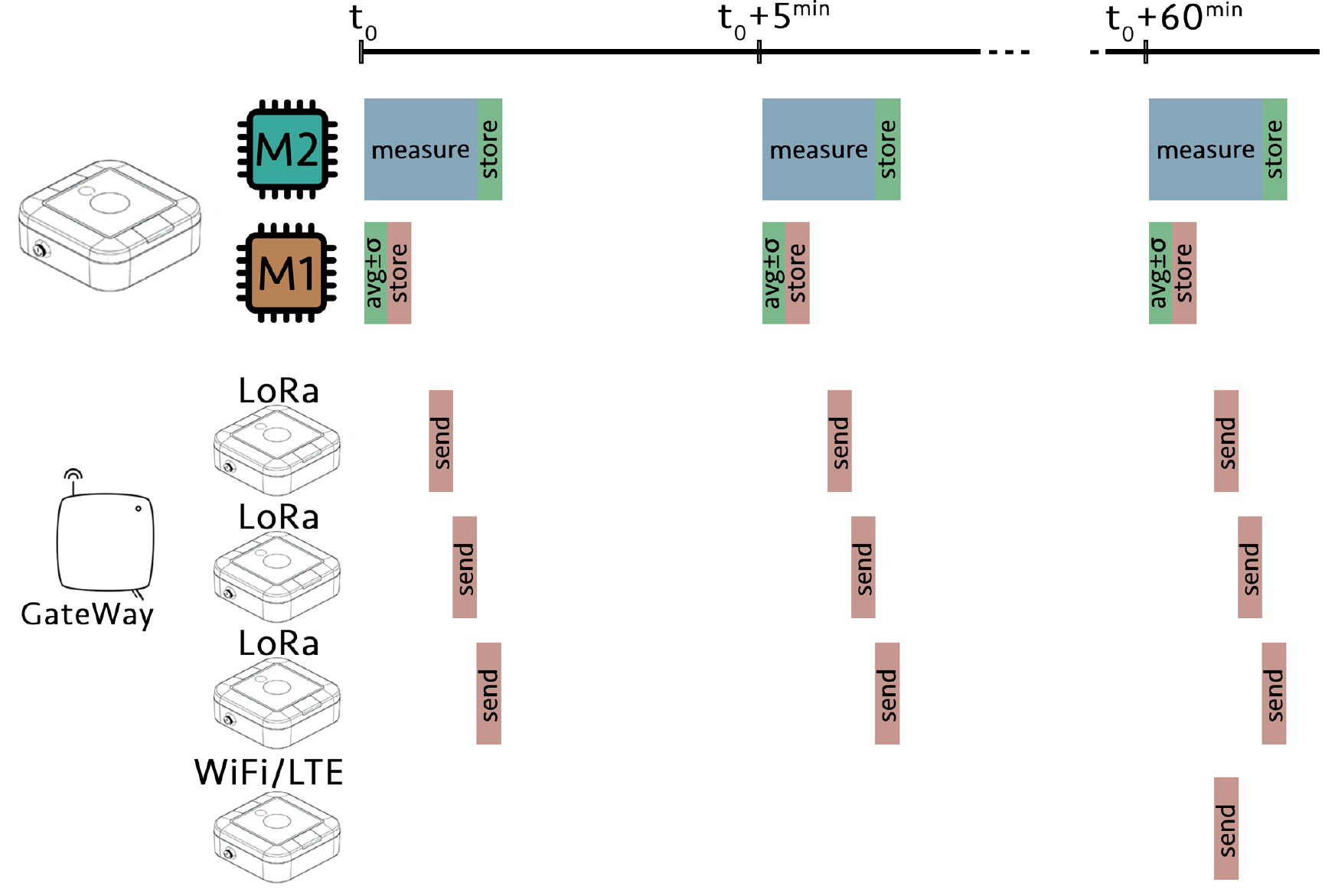}
\caption{Measurement and communication diagram of the SG-WAS photometer. Each device has two microcontrollers: M2 takes ten NSB measurements and stores them temporarily; M1 calculates the average and uncertainty of the data set taken in the previous slot, and~stores them until they are transmitted to the central hub. This process is repeated every 5 min. In~the LoRa version, the~GateWay assigns a 6 s slot every 5 min to each photometer to send its measurements and synchronize its timestamp for the next cycle. The~GateWay sends the measurements of all the devices connected to it to the server. In~the WiFi and LTE-M version, the~measurements are directly sent to the server once every hour and M1 is continuously synchronized with the network. During~the rest of the time, both microcontrollers remain in a deep sleep state, so the battery consumption is extremely low.}
\label{fig:diagram}
\end{figure}

In recent decades, there has been a growing interest in installing sensor networks to monitor 
different variables related to environmental protection \citep{Hart2006,Vairamani2013}, ecology~\cite{Zhang2004, Vera2019}, and volcanology \citep{Awadallah2019}, among~others, in~remote or 
difficult-to-access locations with no infrastructure. These wireless sensor networks (WSN) are 
organized as a series of dispersed nodes with sensing capabilities that collect and send data to 
a centralized hub. All SG-WAS photometers are independent NSB sensors conceived to 
be used as one of these nodes. The~measurement and communication process is shown in
Figure~\ref{fig:diagram}. There are \mbox{two microcontrollers}~incorporated:
\begin{itemize}
    \item M2 is connected to the sensor, takes measurements, and stores them temporarily.
    \item M1 handles timing, storage, data processing, and communication. It is connected to M2 
and the central hub (GateWay in LoRa version or server in WiFi/LTE-M version).
\end{itemize}

To reduce power consumption as much as possible, two independent processes are carried~out:
\begin{enumerate}
    \item Every 5 min, M1 calculates the average and instrumental uncertainty of the ten 
measurements taken by M2 in the previous slot, stores them, and commands M2 to take a new set of measures.
    \item In the corresponding slot, depending on the device version, M1 encrypts, and sends 
the measurements stored in its memory to the central hub. These include the average and
 standard deviation of the ten continuous NSB measurements, ambient and sky temperature, battery 
charge, and communication signal strength.
\end{enumerate}

During the rest of the time, both microcontrollers remain in a deep sleep state, so the battery 
consumption is extremely low. Different versions of SG-WAS depend on the communication 
protocol (WiFi, LoRa, or LTE-M), the~connection to the central hub and how the sending slots are~assigned.

\subsubsection{LoRa~Version}
LoRa \citep{Lora2015} is a patented wireless communication technology commercialized by Semtech 
Corporation since 2012. Signals are modulated in sub-GHz license-free Industrial, Scientific, 
and Medical (ISM) radio bands, making it affordable, deployable worldwide, and interoperable. 
With its ability to reach long distances of up to 15 km in rural areas, provide data rates in the range of 
kilobits per second, and more inexpensive base stations (GateWay), LoRa is a very attractive 
technology for WSN communication in remote~locations.

Although there are some ALOHA-based communication protocols with random access channels---LoRaWAN 
\citep{Lorawan2015} is one of most widely known---that support networks with thousands of
 sensors connected to the same GateWay, networks with densities of less than one photometer per 
square kilometer are not expected \citep{Bara2018}. Therefore, each GateWay is limited to several 
tens of photometers. In~addition, random access channels require data re-sending or active listening 
strategies to ensure the reception of all the packets, which is a power-consuming process. 
Consequently, a~deterministic communication protocol of Time Division Multiple Access (TDMA~\cite{Miao2016}) has been~developed.

When a new photometer is added to the network, a~connection with the GateWay is established 
and a 6 s slot in the 5 min cycle is assigned. Within~these 6 s, data collected 
by the device are sent to the GateWay, which returns a confirmation message and synchronizes the 
time to the next slot. The~photometer will not connect again until then and enters a deep sleep 
state. All data collected by the devices in the network are sent to the server by the GateWay via 
WiFi--MQTT~protocol. 

Assuming a 10\% error in the slot time (a total sending time of 6.6 s), each GateWay can accept 
up to 45 devices using the protocol developed in this work. A~test network of 20 SG-WAS LoRa has 
been successfully established around the Teide Observatory (OT, Canary Islands, Spain).

\subsubsection{WiFi and LTE-M~Version}
The WiFi version incorporates a communication module via the MQTT protocol. The~LTE-M version has a 
2G multiband card and may incorporate an additional antenna. The~slotted communication algorithm
 described above is not used with these versions because devices are directly connected to the 
internet. Data are stored in each device (M1, see \mbox{Figure \ref{fig:diagram}}) and sent to the server every~hour.

\subsection{Autonomous}
\label{sec:Autonomous}
The measurement and communication algorithm described above has been developed to optimize the power
 consumption of the device. It incorporates a solar panel at the top that recharges the internal 
Li-Ion battery under direct sunlight (nominal 5.5 V and 35 mA with solar irradiance of 455 W/m$^{2}$).

\begin{figure}[t]
\includegraphics[width=13.5 cm]{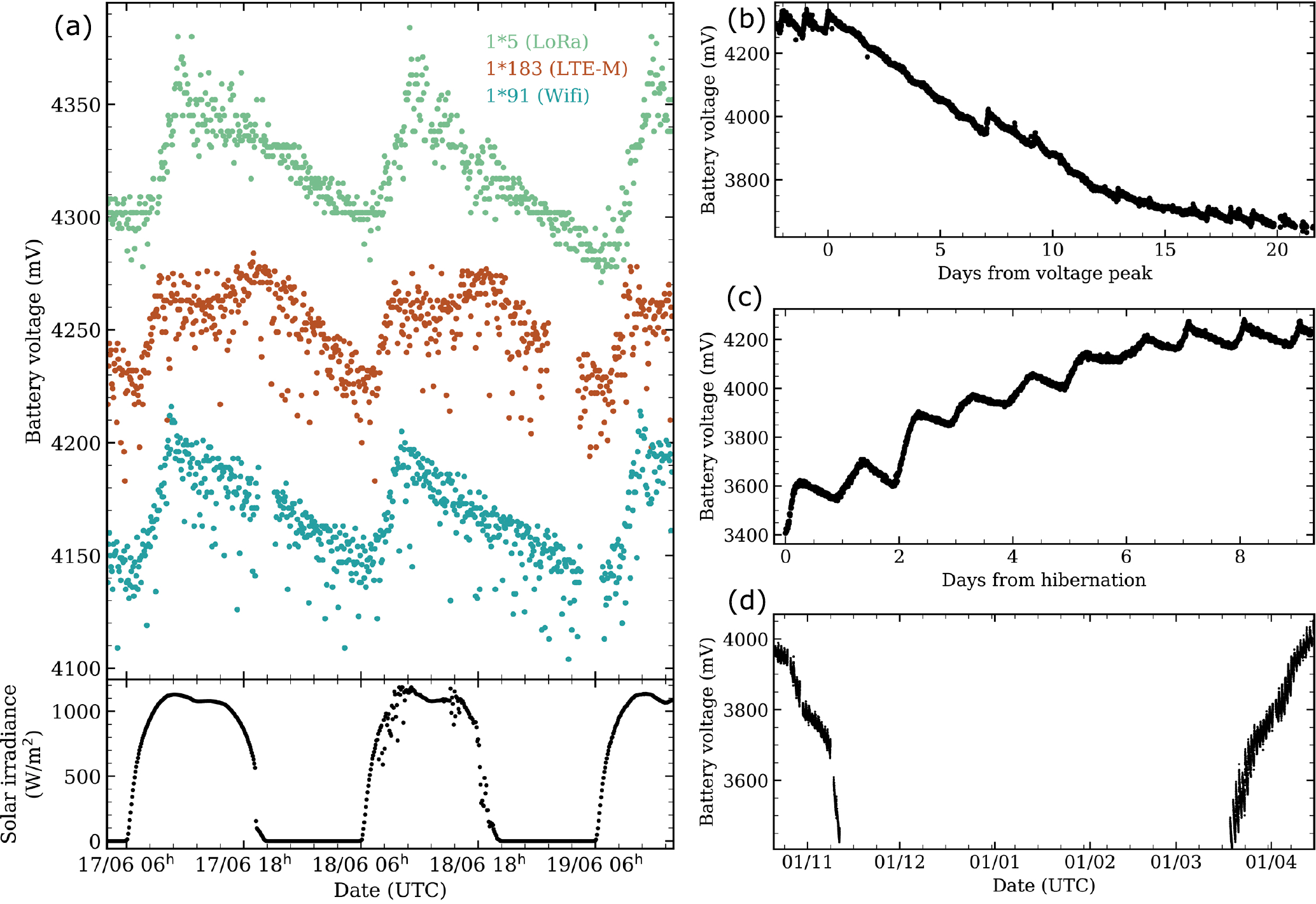}
\caption{(\textbf{a}) Normal charge --discharge cycle for three photometers SG-WAS LoRa (green), LTE-M (brown), 
and WiFi (blue) versions located at the Teide Observatory (OT, Canary Islands, Spain). The~solar 
irradiance, measured by the Stella telescopes weather station, is shown at the bottom; \mbox{(\textbf{b}) discharge} 
curve of an SG-WAS WiFi when not exposed to direct sunlight. It takes more than \mbox{20 days} to stop 
taking measurements uninterruptedly until it reaches the hibernation state; (\textbf{c}) charge curve of an SG-WAS 
once it receives direct sunlight from the hibernation state; (\textbf{d}) voltage curve of the device 1$*$53  
placed in the Valley of Tejeda (Gran Canaria, Spain), which is no longer exposed to direct sunlight 
in mid-November, remains in hibernation for 4 months and is charged in March when it begins to receive 
sunlight again.}
\label{fig:battery}
\end{figure}

Many battery life tests have been performed with the different SG-WAS versions to study their 
autonomy. Figure~\ref{fig:battery}a shows the charge and discharge curve of three devices located at 
the OT, under~a daytime solar irradiance \textasciitilde1000 W/m$^2$. In~all versions, the~charge lost during 
the night is recovered in the first 4 h of the day even in the presence of thin clouds (see a less 
smooth curve on 18 June), at~a rate of 15--20 mV/hr. Once the voltage peak is reached, the~battery 
begins to discharge slowly at a rate of $\sim$2.5 mV/h. During~the self-charging and discharging 
phases, measuring and communication is not interrupted at any~time.

A complete discharge curve is shown in Figure~\ref{fig:battery}b. The~1$*$101 (WiFi) photometer was 
placed inside a telescope dome that was closed during daytime. It took more than \mbox{20 days} for the device to reach 
its battery hibernation point. This checkpoint is introduced to prevent the battery from reaching the 
deep discharge limit and being damaged. Once re-exposed to sunlight, the~device begins to charge and
 send measurements again, reaching the peak voltage just a week later (Figure \ref{fig:battery}c). 
The hibernation phase can be maintained for months; see Figure~\ref{fig:battery}d for the battery 
voltage of photometer 1$*$53 located in the Valley of Tejeda (Gran Canaria, Spain). It is located on
 a rocky point that does not receive sunlight during the winter and part of the autumn, so the 
photometer started to discharge and went into hibernation in mid-November 2020. Four months later, 
in mid-March 2021, it started receiving direct sunlight, sending measurements and recharging again, 
returning to its normal operating state. The~battery stability and autonomy of the device has been 
extensively tested~successfully.

\subsection{Sensor}
\label{sec:Sensor}
TSL237 is a silicon photodiode combined with an electrical current-to-frequency converter on a single 
monolithic CMOS integrated circuit. It was designed by Texas Advanced Optoelectronic Solutions 
\citep{Ams} and is the most widely used sensor in low-cost photometers because of its low price, 
high sensitivity, and accuracy. TSL237 covers a spectral range between 300 nm and 1050 nm with 
maximum sensitivity at 700 nm, which lies between the visible and near infrared. According to the 
manufacturer, it can work between $-$25 ºC and 70 ºC, and  is thermally compensated between 320 nm and 
700 nm. The~output signal is a square wave with a frequency that is linearly proportional to the intensity of 
the light (irradiance) incident on the photodiode in the range between 1 Hz and 1 MHz. A~1.8 mm 
diameter lens is integrated in the optical center and 0.07 mm above the~photodiode.

\subsubsection{Linearity}
\label{linearity}
According to the manufacturer \citep{Ams}, the~sensor linearity is fulfilled in the range between 1 Hz 
and 1 MHz when illuminated with a spectral lamp at 524 nm at ambient temperature. This has been 
verified before with several SQM devices down to 19 mag/arcsec$^2$ \citep{Pravettoni2016}. However, 
measuring in darker places requires studying the literality down to magnitude 22, corresponding to sub-Hz~frequencies.

\begin{figure}[t]
\includegraphics[width=13.5 cm]{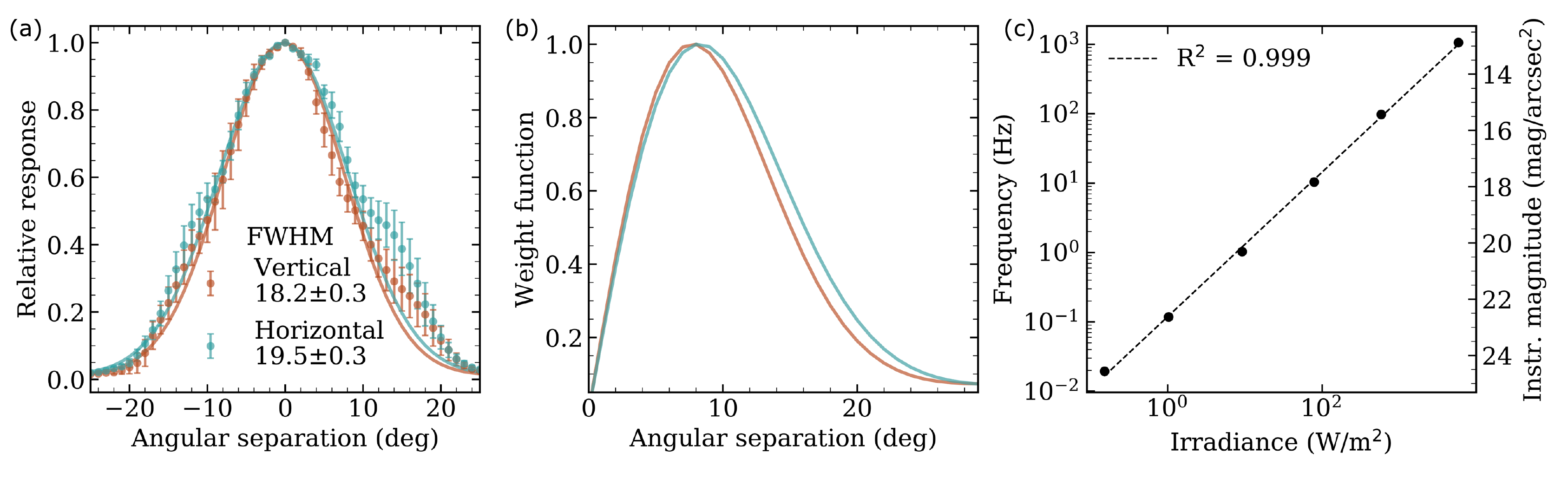}
\caption{(\textbf{a)}): FOV in the vertical (brown, defined in the IR window-thermometer direction) and 
horizontal (blue) axis, obtained in the laboratory by measuring received irradiance for different angles 
with respect to the optical axis. The~FWHM obtained from the Gaussian fit in both axes is included. 
(\textbf{b}): Weight function of each FOV ring over the total flux. (\textbf{c}): Frequency measured by the TSL237 
sensor against received irradiance; its approximate equivalence to magnitudes per square arcsecond is 
shown on the right axis. A~fit is included to show the linearity of the detector to magnitudes 
greater than natural sky darkness.
}
\label{fig:fov}
\end{figure}

\textls[-15]{A TSL237, an~AvaSpec-ULS3648-UA-25 (AVANTES) fiber spectrometer and a HALOSTAR 50W 12V GY6.35 (OSRAM) 
lamp were placed inside a black box. The~intensity and position of the lamp were varied to cover the 
range of 0.01 to 10,000 W/m$^2$ received at the entry of the fiber and the sensor, which are placed 
together, while recording the frequency registered by the sensor. The~collected measurements, including 
the correspondence to instrumental magnitude, are shown in Figure~\ref{fig:fov}(c). Linearity 
is maintained down to 0.01 Hz, equivalent to about 24 mag/arcsec$^2$, with~an $R^2$ very close to 1. 
The TSL237 sensor is therefore linear in the measurement of NSB in both light-polluted and natural dark locations.}

\unskip

\subsubsection{Angular~Response}
\label{sec:angular}
As for the SQM-L and TESS-W, a~low-cost concentrator lens is located before the sensor (following 
the optical path), allowing more light to be collected in a narrower FOV. To~verify the angular response 
of the device in its principal axes, it has been placed on a goniometer aligned---0$^\circ$---with a point 
light source. Positive values indicate a clockwise rotation, being negative otherwise. Measurements have 
been taken from $-45$$^\circ$ to 45$^\circ$ in 1$^\circ$ steps with three different devices to avoid possible
 random errors in the mounting of one of them. The~mean values and their deviation are shown in
Figure~\ref{fig:fov}(a). The~FOV is approximately Gaussian in both axes, with~a full width at half maximum 
(FWHM) of $18.2\pm0.3$ deg in the vertical and $19.5\pm0.3$ deg in the horizontal. The~bumps that stand 
out from the fit are due to stray light reflections at very specific positions in the experimental 
set-up. This FWHM is close to that of SQM-L (~20$^\circ$ \cite{Cinzano2007}) and TESS-W (17$^\circ$ \cite{Zamorano2017}).

Knowing the FOV, the~contribution of each integrated ring, defined from the angular separation with 
respect to the optical axis, to~the total flux measured by the device can be obtained. This is a weight 
function that should be considered when calculating the average radiance at the entrance of the detector 
plane---see, for~example, Equation~(\ref{eq:rad}). The~result on both axes is shown in Figure~\ref{fig:fov}(b). Although~the detector response decreases with increasing angular separation, the~integrated 
sky area is larger and the maximum is located approximately 8--9º from the optical axis. This is particularly
 meaningful in the presence of gradients produced by extended artificial (ALAN) or natural (twilight, 
moonlight, galactic plane, etc.) illumination, but~less for point sources such as stars, whose fluxes 
are added to the rest of the sky in the integrated~ring.

\subsubsection{Transmittance}
A window and optical filter were placed in the entrance plane in the early versions of 
SG-WAS . This is a similar configuration to that of the SQM, with~a Hoya CM-500, and~TESS-W, with~a 
dichroic filter. The~empty gap between the two surfaces is exposed to condensation from any moisture 
that may remain inside the case. Manufacturers of the TESS-W included a heater to keep the internal 
temperature above 10 °C and avoid such condensation. Given the autonomy requirements of the SG-WAS, this
 energy-intensive solution (more power-consuming than the measurement and communication processes) is 
not feasible. The~dichroic$+$window assembly has been replaced, then, by~a single dichroic, which limits 
the spectral response of the detector and ensures proper sealing of the~device.

\begin{figure}[t]
\includegraphics[width=13.5 cm]{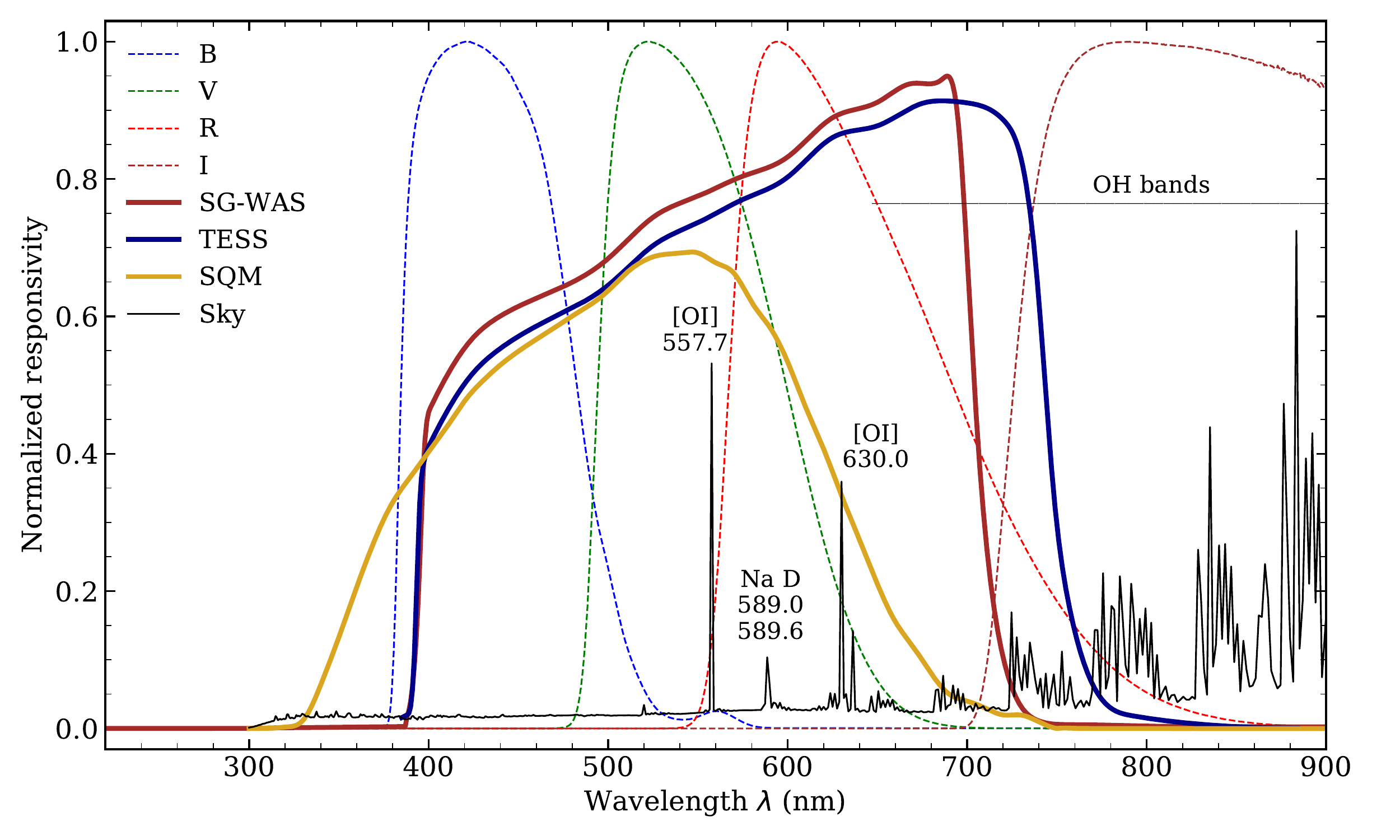}
\caption{Spectral response curve of the SG-WAS. The~sensitivity of the TSL237 sensor is limited to 
the visible range by a dichroic filter, and~the window has been removed to improve the transmittance 
of the optical system. The~transmittance of the TESS-W and SQM-L photometers and Johnson--\mbox{Cousins 
\citep{Bessell1990}} \textit{BVRI} filters has been included as a reference, as~well as the night sky 
spectrum obtained using the SkyCalc tool \citep{Noll2012}, where the brightest airglow lines are labeled. Adapted with permission from \cite{Alarcon2021} ©AAS 2021.}
\label{fig:spectra}
\end{figure}

This is also an opportunity to optimize the transmittance of the device, a~critical consideration if an additional 
color filter is added to the SG-WAS optics (under development). The~window causes an approximately 8\% 
loss in transmittance and the filter, 6\%. If~they were considered as two independent elements, the~ total transmittance loss would be approximately 13\%. Removing the window causes the optical system to become more
 sensitive, as~shown in Figure~\ref{fig:spectra}. The~spectral range of SG-WAS is fitted with 
\textit{BVR} filters, from~400 to \mbox{720 nm}. These filters cover the main optical airglow~lines.

\unskip

\section{Results and~Discussion}
\label{sec:results}
\unskip
\subsection{SG-WAS Astronomical Magnitudes and~Uncertainties}
\label{sec:Sky Brightness}

Following standards in astronomy, brightness is usually expressed in units of magnitude per square 
second as follows:
\begin{equation}
\label{eq:mag}
    m = ZP - 2.5\log_{10}(f-f^{\rm D}),
\end{equation}
where $ZP$ is the laboratory-defined zero point of the photometer using a reference light source, and
 $f$ and $f^{\rm D}$ are the signal and dark frequencies of the sensor, respectively, both measured 
in Hz. To~avoid mistakes, it is more correct to consider the SQM, TESS, and SG-WAS responses as different
 photometric systems. The~conversion factors between photometric systems can be obtained based on the 
kind of spectra of the observed \mbox{object \citep{Bar2019}}. Hereafter, the~brightness in mag/arcsec$^{2}$
 measured in the SG-WAS passbands will be called m$_{\rm SG}$ and Equation \eqref{eq:mag} is transformed into
\begin{equation}
\label{eq:magSG}
    m_{\rm SG} = ZP -2.5\log_{10}(f_{\rm SG}-f_{\rm SG}^{\rm D}).
\end{equation}

\textls[-15]{At typical nighttime low operating temperatures (<20$^\circ$), the~value of the dark frequency, 
$f_{\rm SG}^{\rm D}$, turns out to be negligible in most cases; therefore, the~total error of the 
measurements of the SG-WAS photometer for output frequency  $f_{\rm SG}$, $\delta^2 m_{\rm SG}$, 
is obtained and converted to device-specific brightness units by propagating the expression \eqref{eq:magSG} as follows:}
\begin{equation}
\label{eq:magSGerror}
    \delta^2 m_{\rm SG} =\delta_{ZP}^2 +
    ~\left(\frac{2.5\log_{10}(e)}{f_{\rm SG}}~\delta_{f_{\rm SG}} \right)^2,
\end{equation}
where
\begin{equation}
\label{eq:magSGerror1}
     \delta_{ZP} = \rm Calibration ~Error
\end{equation}
\begin{equation}
\label{eq:magSGerror2}
    \delta m_{\rm SG}^{ins} =~\frac{2.5\log_{10}(e)}{f_{\rm SG}}~\delta_{f_{\rm SG}}
     = \rm Instrumental ~Error.
\end{equation}
Both calibration and instrumental errors must be taken into account when determining 
the total uncertainty of the~measurements.

\subsection{Sky Integrating Sphere (SIS) Calibration Method}
\label{sec:Calibration}
The determination of the $ZP$ is the key issue in the reproducibility of measurements and 
consistency between different devices. Starting from the expression \eqref{eq:magSG}, and considering 
that the measurement of the reference device $m_{\rm SG_{ref}}$ is expressed in the photometric 
absolute (AB) magnitude system, the~difference between this and a simultaneous measurement of another 
device will be given by
\begin{equation}
    m_{\rm SG} - m_{\rm SG_{\rm ref}} = ZP -2.5\log_{10}\left(\frac{f_{\rm SG}-f_{\rm SG}^{\rm D}}{f_{\rm SG_{\rm ref}}-f_{\rm SG_{\rm ref}}^{\rm D}}\right)\label{eq:mf}
\end{equation}

In the comparison calibration process, we are interested in establishing the $ZP$ value from 
the magnitude difference. Under~the same illumination conditions, two correctly calibrated 
photometers, i.e., with the $ZP$ well determined, should measure the same magnitude within the 
uncertainty range, so the ratio between $(f_{\rm SG}-f_{\rm SG}^{\rm D})/(f_{\rm SG_{\rm ref}}-f_{\rm SG_{\rm ref}}^{\rm D})$ 
is constant. The~relationship between the frequency measured by the TSL237 sensor and the 
spectrally weighted and FOV-averaged radiance $L$ at the entrance plane of the detector is given by
\begin{equation}
    f = K \int_{0}^{\infty} T(\lambda)\left[\int_{\Omega}P(\boldsymbol{\omega}) L_\lambda(\boldsymbol{\omega})d^2\boldsymbol{\omega}\right] d\lambda + f^D \label{eq:rad}
\end{equation}
where $L_\lambda(\boldsymbol{\omega})$ is the spectral radiance of the incident light field
 along the direction specified by the angular vector $\boldsymbol{\omega}$, $P(\boldsymbol{\omega})$ 
is the weight function describing the FOV of the device, normalized such that 
$\int_{\Omega}P(\boldsymbol{\omega})d^2\boldsymbol{\omega} = 1$, $T(\lambda)$ is 
the spectral response of the device and $K$ a constant that provides the absolute 
link between the converter output frequency and incident radiance \citep{Bar2019}. 
Under angularly uniform radiance illumination (i):
\begin{equation}
    f = K \int_{0}^{\infty} T(\lambda) L_\lambda d\lambda + f^D
\end{equation}
which is device-independent only if (ii) $f^D$, which varies with temperature, is the 
same, i.e.,~$f_{\rm SG}^{\rm D} = f_{\rm SG_{\rm ref}}^{\rm D}$, or~negligible; (iii) there 
is no significant change for both devices in the spectral response $T(\lambda)$, bearing in mind
that the incident radiance has the same spectral distribution, i.e.,
 $L_\lambda^{\rm SG} = L_\lambda^{\rm SG_{\rm ref}}) $. These three conditions must be 
fulfilled to obtain the $ZP$ from the difference of simultaneous measurements of two~photometers.

A well-established procedure is to use a stable light source to illuminate a homogeneous 
surface and take simultaneous measurements between a reference photometer and the one to
 be calibrated, thereby establishing the ZP from the differences. In~both the SQM-L and TESS-W calibrations, 
an integrating sphere is used to ensure homogeneous irradiance. In~the first case, a~broadband 
spectral lamp is used, with~increasing irradiance in the wavelength range of 350 to 500 nm that remains almost 
constant thereafter \citep{Cinzano2005} and, in~the second one, a~LED with a spectral response 
centered at 596 nm and an FWHM of \mbox{14 nm \citep{PosterZamorano2017}}. This system presents several 
difficulties in ensuring reproducibility and usually requires a 1-by-1~calibration.

On the one hand, a~dependence on the experimental setup is introduced that may lead to 
increased measurement uncertainty (as reported by~\cite{Pravettoni2016}) due, for~example, to~imprecise positioning or poor screening of the background stray light breaking the angular 
uniformity (i). Moreover, it requires careful control of the temperature at which the 
measurements are taken, as~this may not only affect the emission spectrum of the lamp but
 also the dark frequency, which may not be negligible at working temperature (ii).

On the other hand, it is known that the spectral response of different photometers is not 
the same, so condition (iii) is not always satisfied. In~the extreme case, it is not correct 
to obtain the ZP by comparing simultaneous measurements of a TESS-W and an SQM-L because their 
spectral response is very different and hence also the frequency measured, even under the 
same illumination conditions. This effect also occurs, although~to a lesser extent, in~
photometers of the same type. When calibrating with a narrow source, as~in the case of the 
TESS-W, it is assumed that the difference in magnitudes obtained in the spectral range of the 
lamp can be extrapolated to the full spectral response of the device, thus introducing an undesired
 and unknown systematic error. To~reduce this systematic error in night sky brightness 
measurements, a~light source with a spectral distribution similar to that of the night sky may 
be much more~appropriate.

In this work, we have developed a novel calibration method based on simultaneous measurements
 of the night sky that can be used in any type of NSB device. The~night sky is likely to behave 
as an integrating sphere if there is no light gradient in the photometer FOV.  Therefore, it is 
necessary to avoid localized sources of both artificial (ALAN) and natural light emission
 (astronomical components such as the Moon or the Galaxy). All SG-WAS photometers are calibrated 
on zenith observations taken from the OT, where light pollution is minimal \citep{Alarcon2021} and 
condition (i) satisfied (further details on the calibration procedure are omitted because of 
a patent-pending process). When taking simultaneous nighttime measurements, not only is the 
temperature the same in all devices, but~also, at~an altitude of 2500 m, it does not get high 
enough for the dark frequency to be significant. In~addition, many photometers (currently up to 100 devices at the OT) can be calibrated 
at the same time as there is space available on the optical~table.

The most meaningful point of this method is that the calibration measurements are taken with the 
same spectrum of the night sky. {This allows the systematic error to be very well constrained to the absolute calibration of the reference device.} The difference between the measurements {is thought to come mainly from the small variations} in the spectral response of the photometer, resulting in a variation of the magnitude value with the changes
that take place in the night sky spectrum over~time. 

\begin{figure}[t]
\includegraphics[width=13.5cm]{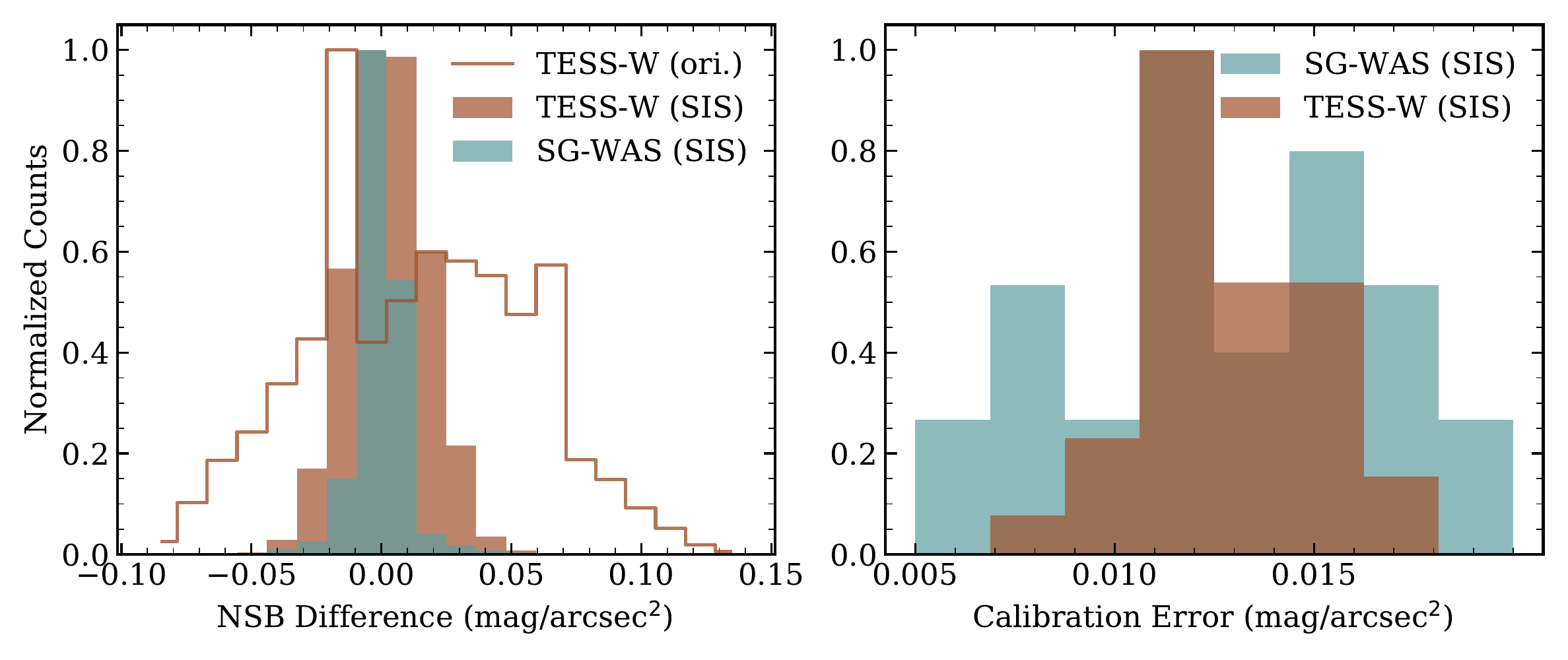}
\caption{(\textbf{a}): Distribution of differences in simultaneous NSB measurements between pairs of 
TESS-W (brown) and SG-WAS (blue) photometers after calibrating them using the Sky Integrating 
Sphere (SIS) method. The~distribution of TESS-W photometer pairs with the original calibration 
(ori.) given by the manufacturer is also shown unfilled. (\textbf{b}): Distribution of calibration errors for both photometers obtained with the SIS method.}
\label{fig:SIS}
\end{figure}

\begin{figure}[t]
\includegraphics[width=13.5cm]{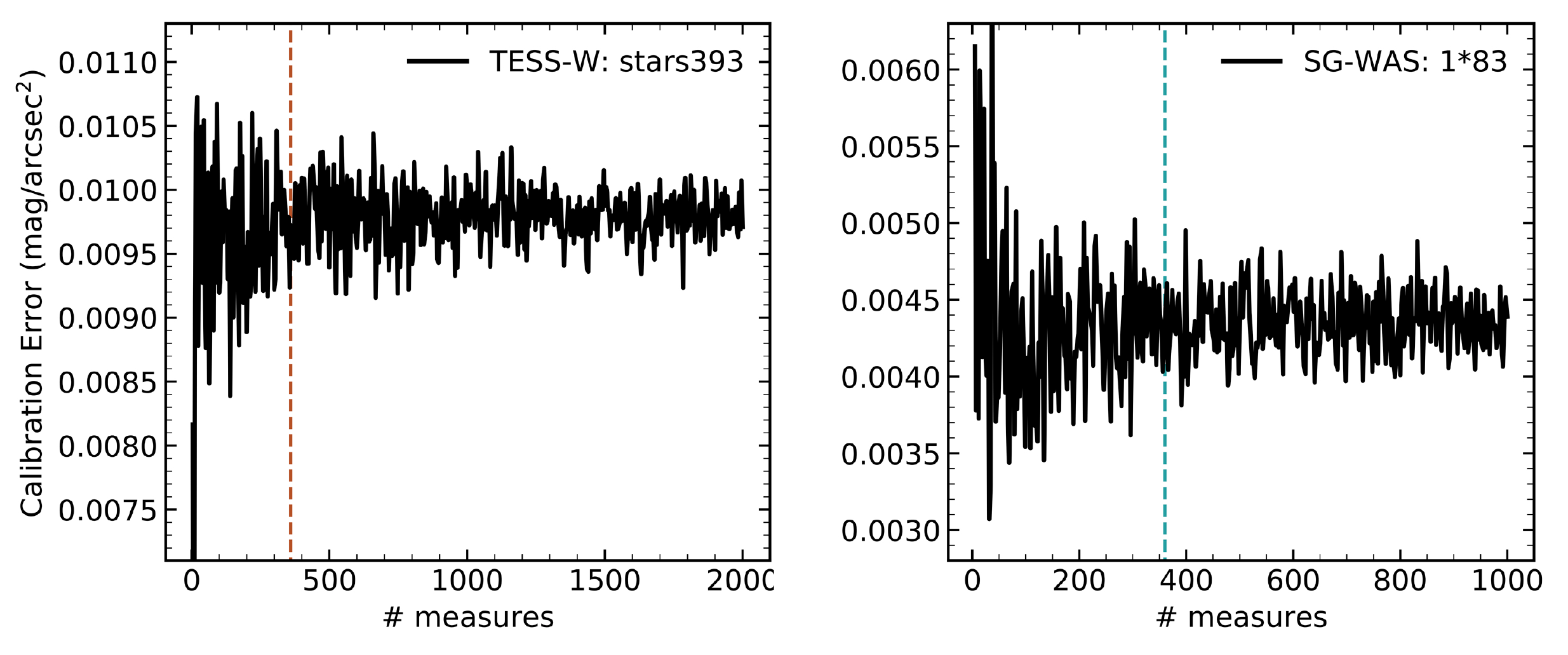}
\caption{Calibration error
 obtained with the SIS method for a TESS-W photometer (\textbf{a}) and an
 SG-WAS (\textbf{b}), as~a function of the number of measurements taken randomly. It is considered 
that the lower limit is reached from 360 onward (dashed vertical lines), corresponding to 6 and 
30 h of night sky measurements, respectively.\label{fig:measures}}
\end{figure}

\subsection{SG-WAS Calibration~Error}
\label{sec:cal-error}

The SIS method was used to obtain the ZP of 10 TESS-W and 21 SG-WAS photometers. The~differences 
between simultaneous zenith measurements taken between pairs of photometers under dark conditions 
(following the definition described in~\cite{Alarcon2021}) for the period 2021 January 18--May 26  
 and for TESS-W and between 2021 April 14--June 10  for SG-WAS are shown in Figure~\ref{fig:SIS}(a). Both distributions are centered at 0, as~expected, and~have a standard deviation of 0.01 
mag/arcsec$^2$ in the case of SG-WAS and 0.02 mag/arcsec$^2$ for TESS-W. These values are greater 
than the combination of instrumental errors (more detail in the next section) and are due to the 
effect of variations in the night sky spectrum captured by a slightly different spectral response 
in the devices. The~uncertainty in the determination of the ZP by the SIS method does not exceed 
0.02 mag/arcsec$^2$, as~it is shown in  Figure~\ref{fig:SIS}(b). This is a reduction of 
more than a factor 2 of the 0.044 mag/arcsec$^2$ reported as the calibration error in the TESS-W 
\citep{PosterZamorano2017, Bar2019} and a factor of 5 of the SQM \citep{Pravettoni2016}.

The original calibration of the TESS-W, which is made by the manufacturer in the Laboratory for
 Scientific Advanced Instrumentation (Laboratorio de Instrumentación Científica Avanzada, LICA) 
of Universidad Complutense de Madrid (UCM, Spain), presents a more scattered distribution in the 
differences, with~a standard deviation of 0.042 mag/arcsec$^2$ (see the unfilled distribution in 
Figure~\ref{fig:SIS}). 

Figure~\ref{fig:measures} shows the variation of the calibration error for a TESS-W and an SG-WAS 
as a function of the number of measurements considered, taken randomly from the whole set. The~
choice of these particular devices is arbitrary, and~their reproducibility has been checked for all 
other available photometers. From~360 measurements onward, the~ZP determination is considered to 
have reached its accuracy limit and the calibration error cannot be further reduced. Given the 
temporal resolution of the photometers, this is equivalent to 6 h of measurements in the TESS-W 
and 30 h in the SG-WAS. Although~the time available to take such measurements depends on 
different astronomical and atmospheric components, it has been verified that this interval is 
easily achievable in one week at the OT, regardless of the time of year. The~SIS method presented 
in this work not only reduce the calibration uncertainty but also allows a massive calibration 
to be performed --currently up to \mbox{100 devices}-- by taking simultaneous measurements for no longer 
than one clear~week.

\subsection{SG-WAS Instrumental~Error}
\begin{figure}[t]
\includegraphics[width=13.5 cm]{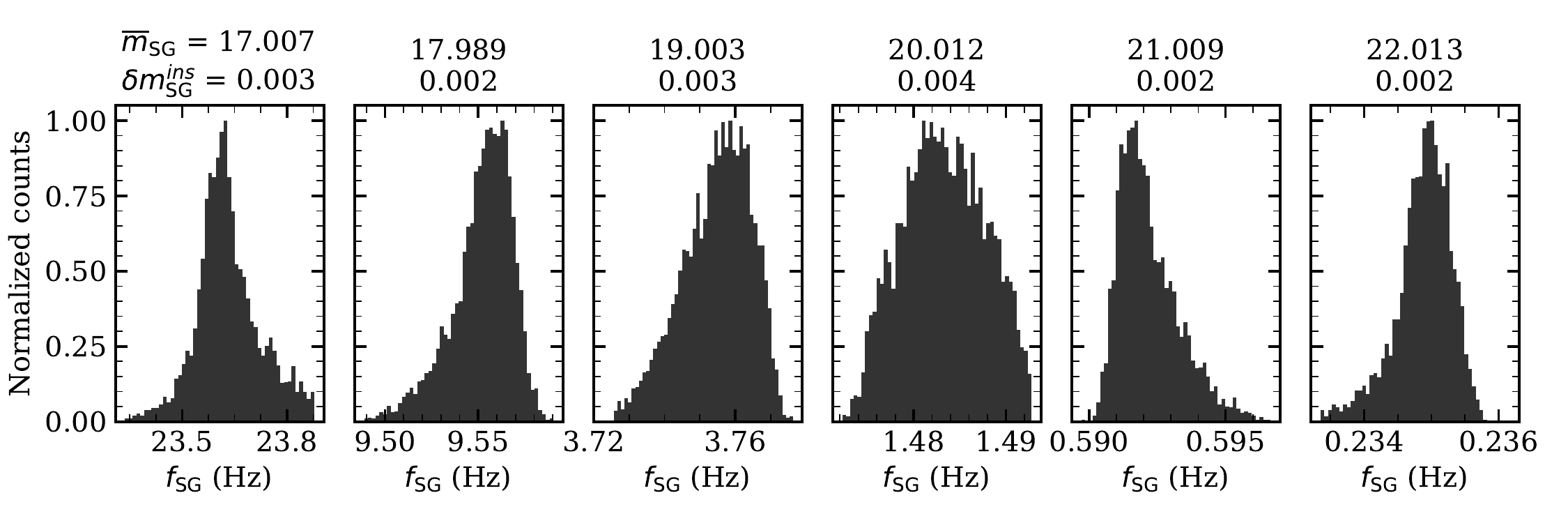}
\caption{Distribution of 5000 laboratory frequency measurements with light intensities for 
$m_{\rm SG}$ values included between 17 and 22 mag/arcsec$^2$ range in steps of 1 mag/arcsec$^2$. 
The standard deviation in magnitude units, corresponding to the instrumental uncertainty, is included on top.}
\label{fig:instr_SG}
\end{figure}

A statistical study of 
the data has been carried out to characterize the instrumental error of the system (sensor$+$optics). 
For this purpose, 5000 laboratory frequency measurements were taken for 
magnitude values between 17 and 22 mag/arcsec$^2$ (as expected in the field concerned) in steps of 
1 mag/arcsec$^2$. The~optical setup is similar to that used in previous sections. For~each set 
of 5000 measurements, a~histogram has been obtained and represented in Figure~\ref{fig:instr_SG}. 
Statistical parameters are shown in Table~\ref{tab:instr_SG}.  

\begin{specialtable}[t] 
\caption{Frequency standard deviation values for each distribution. Following Equation 
\eqref{eq:magSGerror2}, the~SG-WAS instrumental magnitude, $\delta m_{\rm SG}^{ins}$, was also calculated.} 
\label{tab:instr_SG}
\begin{tabular}{@{\hspace{5mm}}c@{\hspace{18mm}}c@{\hspace{18mm}}c@{\hspace{17mm}}c}
\toprule
\textbf{Frecuency} & \textbf{Magnitude}	& \boldmath{\textbf{$ \delta_{f_{\rm SG}} = \sigma_{f_{\rm SG}}  $}}	& \boldmath{\textbf{ $\delta m_{\rm SG}^{ins}$}}  \\
\textbf{(Hz)} & \textbf{(mag/arcsec$^2$)} & \textbf{(Hz)} & \textbf{(mag/arcsec$^2$)} \\
\midrule
9.5563	&	17.989 	&	0.0134  & 	0.002 \\
3.7564	&	19.003 	&	0.0100  & 	0.003 \\
1.4833	&	20.012 	&	0.0053  & 	0.004 \\
0.5918	&	21.009 	&	0.0009  & 	0.002 \\
0.2350	&	22.013 	&	0.0004  & 	0.002 \\
\bottomrule
\end{tabular}
\end{specialtable}

The calculated standard deviation values for each distribution magnitude is lower than  
0.004 mag/arcsec$^2$ (see Table~\ref{tab:instr_SG}). This experiment therefore shows 
that the instrumental error of the SG-WAS photometer is lower than a few milli-magnitudes. 
Taking into account that this laboratory error calculation  can still be affected by other 
factors such as the power supply error, the~stability of the lamp used, the~design of the 
experiment (background stray light), electromagnetic interference, etc., it should be 
considered an upper limit to the instrumental~error. 

The new measurement method developed for the SG-WAS allows us to obtain the average of ten 
continuous measurements of NSB and also its standard deviation. The~time taken by the 
device to take a measurement is inversely proportional to the frequency. Following Table~\ref{tab:instr_SG}, it takes approximately 4.5 s to take one measurement at magnitude 22, 1.7 s at 
magnitude 21, and so on. In~total, ten continuous measurements at magnitude 22 are taken in about 
45 s. As the possible variations in sky brightness produced by airglow are not 
measurable on this time scale \citep{Alarcon2021}, the~standard deviation of these 
measurements corresponds only to the instrumental error, which is provided along with the 
NSB every \mbox{5 min} in the~SG-WAS.

\subsection{Night Sky Brightness~Measurements}

\begin{figure}[t]
\includegraphics[width=13.5cm]{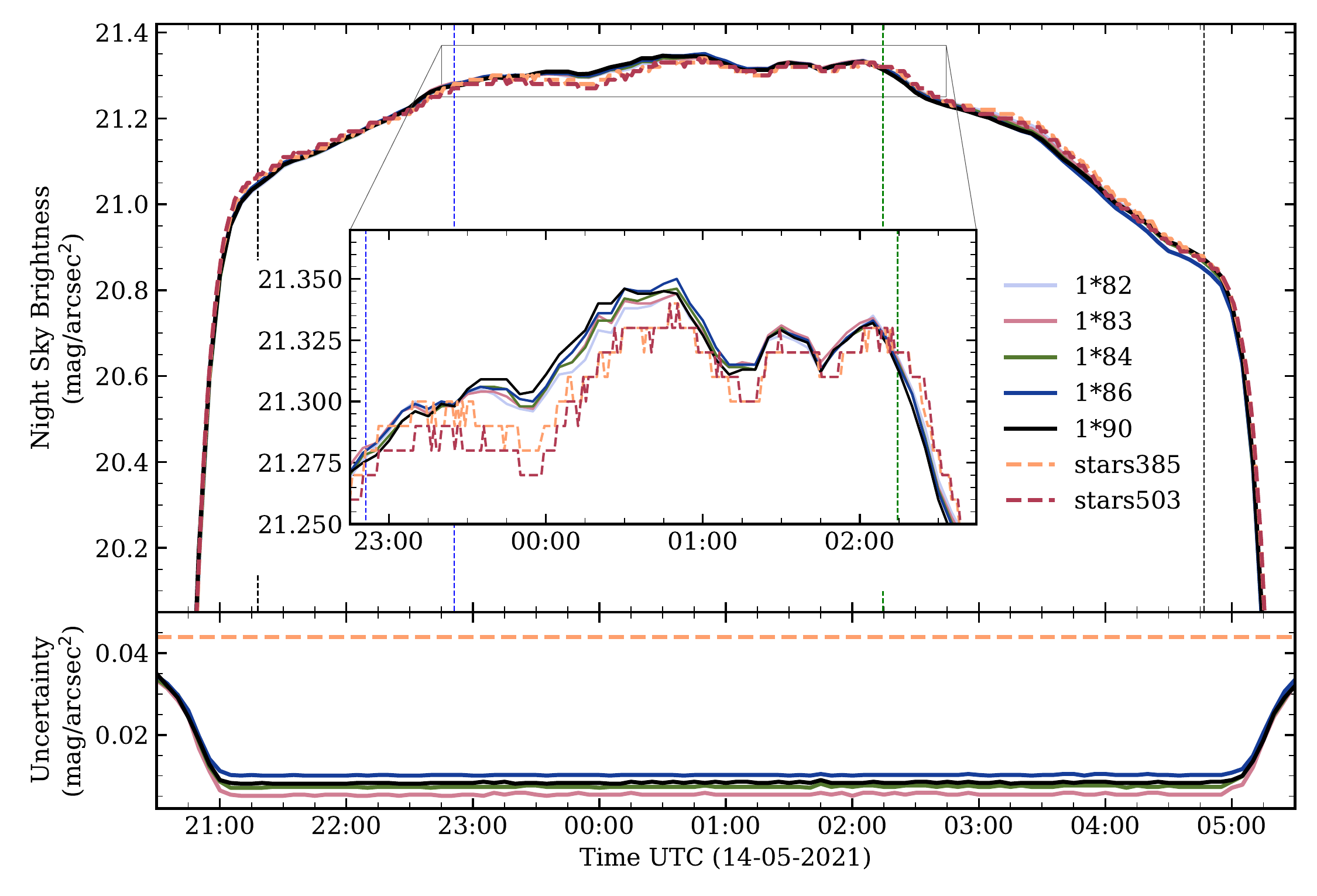}
\caption{NSB curve of five SG-WAS (1$*$X) and two TESS-W (starsX) photometers taken on 2021 May 14 
from Teide Observatory (Canary Islands, Spain). The~vertical lines indicate the end (\textbf{left}) 
and beginning (\textbf{right}) of astronomical twilight (black), the~leaving of the zodiacal light 
(blue) and the entrance of the Galaxy (green) in the FOV of the device. The~inner graph shows 
the dark \mbox{period \citep{Alarcon2021}} enlarged. The~SG-WAS curves are smoother than those of TESS-W as a consequence of the 10-measurement averaging process. The~lower plot shows the 
total uncertainty in the measurements, including the instrumental error---calculated for each 
SG-WAS measurement and its upper bound of 0.004 mag/arcsec$^2$ for TESS-W---and the calibration
 error---obtained by the SIS method for SG-WAS and 0.044 mag/arcsec$^2$ for TESS-W.}
\label{fig:result}
\end{figure}

More than one-hundred SG-WAS photometers were installed at the OT during 2021 to test their resistance 
to extreme weather stress conditions (snow, ice, strong wind, high solar radiation, among~others)
 and the reliability of their NSB measurements. Figure~\ref{fig:result} shows the zenithal 
brightness curve on 2021 May 14 for five SG-WAS and two TESS-W placed together. In~the inner plot, 
the dark and clear period (as defined by \citep{Alarcon2021}) is shown enlarged. Note that the 
SG-WAS curves are smoother than those of TESS-W, which is a result of the averaging process. 
Variations associated with airglow are clearly visible in both devices, although~some differences 
between them are appreciable. This is a consequence of the slight difference in the spectral 
response of both devices.  The~SG-WAS curves are very similar to each other, with~their difference 
being less than one-hundredth of a~magnitude.

The lower plot shows the total uncertainty in the measurements, including the instrumental 
error (calculated for each SG-WAS measurement and its upper bound of \mbox{0.004 mag/arcsec$^2$} for 
TESS-W) and the calibration error (obtained by the SIS method for SG-WAS and 0.044 mag/arcsec$^2$ 
for TESS-W). The~uncertainties of the SG-WAS measurements are below 0.02 throughout the night, except~
at twilight, when the rapid change in sky brightness increases the standard deviation of the ten
 measurements. These curves have been found to be very nearly the same for all other SG-WAS photometers, 
regardless of version and~series.
 
\subsection{Light Pollution Laboratory: IoT-EELab} 
\label{subsec:IOT-EELab}
The measurements collected by the EELabs and STARS4ALL networks are openly available in real-time at the IoT-EELab website\footnote{\url{data.eelabs.eu}}. It is an interactive dashboard that collects, controls, and analyses data from hundreds of internet-connected {NSB sensors and several} all-sky cameras. There are more than 170 million entries available containing brightness data 
collected for more than 500 photometers (400 TESS and 100 SG-WAS).

IoT-EELab includes a simple reading page with NSB data from a single location/single date with 
calculation of the lunar phase, mean and median values of NSB during the astronomical night, slopes of 
brightness curves in different time intervals, sunset and sunrise hours, and~percentage of cloud cover. 
Furthermore, there are statistical tools to obtain histograms, ``hourglass'' diagrams, ``jellyﬁsh'' 
plots, and~monthly and annual NSB trends, among~other features. All raw and filtered data can be downloaded in standard format. IoT-EELab provides a unique opportunity for researchers to experiment with new and existing light pollution~datasets.

\section{Conclusions}\label{sec:conclusion}
In this work, the~characterization of the new SG-WAS, a~low-cost NSB photometer based on the TSL237 
sensor (similar to SQM and TESS), has been carried out. The~main conclusions are summarized as~follows:
\begin{itemize}
    \item SG-WAS is the first wireless NSB sensor to communicate via LoRa, WiFI, or LTE-M and be powered
 by solar energy
 . A~measurement every 5 min is taken continuously and sent using a Time Division 
Multiple Access algorithm to avoid packet collision.
    \item The device recharges to peak voltage in just 4 h with direct solar irradiance after a 
full night of operation. It can stay up to 20 days in darkness while taking measurements, subsequently
 remain in hibernation for months before returning to operation once it is illuminated again.
    \item Its optical design is very similar to the TESS-W, with~a FOV approximately 19$^\circ$ (FWHM) and a
 slightly less red spectral range, from~400 to 720 nm. The~window has been removed to prevent 
condensation on the dichroic and increase transmittance.
    \item A new Sky Integrating Sphere (SIS) calibration method (patent-pending) has been designed and 
demonstrated to achieve calibration errors in TESS-W and SG-WAS below 0.02 mag/arcsec$^2$.
    \item The robustness to adverse weather conditions and the stability of its measurements have been 
demonstrated in field tests. Taking the average and uncertainty of ten continuous measurements makes the NSB 
curves smoother and avoids spikes. Differences between simultaneous measurements of several SG-WAS photometers have 
been found to have a standard deviation of 0.01 mag/arcsec$^2$, several times smaller than their predecessors.
\end{itemize}

\vspace{6pt} 



\authorcontributions{M.R.A., M.P.-S., and M.S.-R. conducted the design of the experiments, the~interpretation 
of the results and the organization and writing of the manuscript. M.R.A., M.S.-R., and S.L.-P. developed the SIS
calibration method and IoT-EELabs tool. The~measurements taken in the laboratory of Sieltec Canarias SL
 were performed by M.P.-S., M.M., and C.L. All authors have read and agreed to the published version of the~manuscript.}

\funding{This research was supported by EELabs a project funded by the European Union INTERREG V-A MAC 
2014--2020. This work has been supported by the STARS4ALL Foundation, which maintains the data infrastructure 
for the TESS photometer network. Sieltec Canarias S.L. used its own resources, in~addition to a grant from 
the Ministry of Economy, Knowledge and Employment, Canary Islands Government, under~PILA 157G0042, 85\% 
co-financed by the ERDF within the 2014--2020 Operational Program and PILA 207G0385 for investment projects 
of SMEs in the Canary~Islands.}

\institutionalreview{Not applicable.}

\informedconsent{Not applicable.}

\dataavailability{Access and download of the data used is openly available at \url{data.eelabs.eu}.}

\acknowledgments{We are grateful to Ramon Negrillo Valdivia (Software Developer, Sieltec Canarias S.L.) 
and Patricia Hernández Rodríguez (Electronics Engineer, Sieltec Canarias S.L.) for their contributions 
to the development of the~SG-WAS.}

\conflictsofinterest{MSR and SLP are members of the EU EELabs project SG-WAS development team. 
The funders had no role in the design of the study; in the collection, analyses, or~interpretation of 
data; in the writing of the manuscript; or~in the decision to publish the~results.}

\end{paracol}

\reftitle{References}


\begin{thebibliography}{999}
\bibitem[H{\"o}lker \em{et~al.}(2010)H{\"o}lker, Wolter, Perkin, Tockner,
  et~al.]{Holker2010}
H{\"o}lker, F.; Wolter, C.; Perkin, E.K.; Tockner, K.
\newblock Light pollution as a biodiversity threat.
\newblock {\em Trends Ecol. Evol.} {\bf 2010}, {\em 25},~681--682.

\bibitem[Gaston \em{et~al.}(2013)Gaston, Bennie, Davies, and
  Hopkins]{Gaston2013}
Gaston, K.J.; Bennie, J.; Davies, T.W.; Hopkins, J.
\newblock The ecological impacts of nighttime light pollution: A mechanistic
  appraisal.
\newblock {\em Biol. Rev.} {\bf 2013}, {\em 88},~912--927.

\bibitem[Bennie \em{et~al.}(2016)Bennie, Davies, Cruse, and Gaston]{Bennie2016}
Bennie, J.; Davies, T.W.; Cruse, D.; Gaston, K.J.
\newblock Ecological effects of artificial light at night on wild plants.
\newblock {\em J. Ecol.} {\bf 2016}, {\em 104},~611--620.

\bibitem[Owens and Lewis(2018)]{Owens2018}
Owens, A.C.; Lewis, S.M.
\newblock The impact of artificial light at night on nocturnal insects: A
  review and synthesis.
\newblock {\em Ecol. Evol.} {\bf 2018}, {\em 8},~11337--11358.

\bibitem[Falchi \em{et~al.}(2016)Falchi, Cinzano, Duriscoe, Kyba, Elvidge,
  Baugh, Portnov, Rybnikova, and Furgoni]{Falchi2016}
Falchi, F.; Cinzano, P.; Duriscoe, D.; Kyba, C.C.M.; Elvidge, C.D.; Baugh, K.;
  Portnov, B.A.; Rybnikova, N.A.; Furgoni, R.
\newblock The new world atlas of artificial night sky brightness.
\newblock {\em Sci. Adv.} {\bf 2016}, {\em 2},~e1600377, doi:\href{https://www.doi.org/10.1126/sciadv.1600377}{10.1126/sciadv.1600377}.

\bibitem[Stark \em{et~al.}(2011)Stark, Brown, Wong, Stutz, Elvidge, Pollack,
  Ryerson, Dube, Wagner, and Parrish]{Stark2011}
Stark, H.; Brown, S.S.; Wong, K.W.; Stutz, J.; Elvidge, C.D.; Pollack, I.B.;
  Ryerson, T.B.; Dube, W.P.; Wagner, N.L.; Parrish, D.D.
\newblock City lights and urban air.
\newblock {\em Nat. Geosci.} {\bf 2011}, {\em 4},~730--731, doi:\href{https://www.doi.org/10.1038/ngeo1300}{10.1038/ngeo1300}.





\bibitem[de~Miguel \em{et~al.}(2020)de~Miguel, Kyba, Zamorano, Gallego, and
  Gaston]{SanchezdeMiguel2020}
de~Miguel, A.S.; Kyba, C.C.M.; Zamorano, J.; Gallego, J.; Gaston, K.J.
\newblock The nature of the diffuse light near cities detected in nighttime
  satellite imagery.
\newblock {\em Sci. Rep.} {\bf 2020}, {\em 10}, doi:\href{https://www.doi.org/10.1038/s41598-020-64673-2}{10.1038/s41598-020-64673-2}.

\bibitem[Jiang \em{et~al.}(2018)Jiang, He, Long, Guo, Yin, Leng, Liu, and
  Wang]{Jiang2018}
Jiang, W.; He, G.; Long, T.; Guo, H.; Yin, R.; Leng, W.; Liu, H.; Wang, G.
\newblock Potentiality of Using Luojia 1-01 Nighttime Light Imagery to
  Investigate Artificial Light Pollution.
\newblock {\em Sensors} {\bf 2018}, {\em 18},~2900, doi:\href{https://www.doi.org/10.3390/s18092900}{10.3390/s18092900}.

\bibitem[Bar{\'{a}}(2018)]{Bara2018}
Bar{\'{a}}, S.
\newblock {Characterizing the zenithal night sky brightness in large
  territories: How many samples per square kilometre are needed?} 
\newblock {\em Mon. Not. R. Astron. Soc.} {\bf 2018
},
  {\em 473},~4164--4173, doi:\href{https://www.doi.org/10.1093/mnras/stx2571}{10.1093/mnras/stx2571}.

\bibitem[{Alarcon} \em{et~al.}(2021){Alarcon}, {Serra-Ricart}, {Lemes-Perera},
  and {Mallorqu{\'\i}n}]{Alarcon2021}
{Alarcon}, M.R.; {Serra-Ricart}, M.; {Lemes-Perera}, S.; {Mallorqu{\'\i}n}, M.
\newblock {Natural Night Sky Brightness during Solar Minimum}.
\newblock {\em Astron. J.} {\bf 2021}, {\em 162},~25, doi:\href{https://www.doi.org/10.3847/1538-3881/abfdaa}{10.3847/1538-3881/abfdaa}.

\bibitem[{Aub{\'e}} \em{et~al.}(2020){Aub{\'e}}, {Simoneau},
  {Mu{\~n}oz-Tu{\~n}{\'o}n}, {D{\'\i}az-Castro}, and {Serra-Ricart}]{Aube2020}
{Aub{\'e}}, M.; {Simoneau}, A.; {Mu{\~n}oz-Tu{\~n}{\'o}n}, C.;
  {D{\'\i}az-Castro}, J.; {Serra-Ricart}, M.
\newblock {Restoring the night sky darkness at Observatorio del Teide: First
  application of the model Illumina version 2}.
\newblock {\em Mon. Not. R. Astron. Soc.} {\bf 2020}, {\em 497},~2501--2516, doi:\href{https://www.doi.org/10.1093/mnras/staa2113}{10.1093/mnras/staa2113}.

\bibitem[Cinzano(2007)]{Cinzano2007}
Cinzano, P.
\newblock \emph{Report on Sky Quality Meter, Version L}. {\bf 2007}. \href{http://unihedron.com/projects/sqm-l/sqmreport2.pdf}{Available online}, accessed on 28-06-2021.

\bibitem[Zamorano \em{et~al.}(2017)Zamorano, Garc{\'{\i}}a, Tapia, de~Miguel,
  Pascual, and Gallego]{Zamorano2017}
Zamorano, J.; Garc{\'{\i}}a, C.; Tapia, C.; de~Miguel, A.S.; Pascual, S.;
  Gallego, J.
\newblock {STARS}4ALL Night Sky Brightness Photometer.
\newblock {\em Int. J. Sustain. Light.} {\bf 2017}, {\em
  18},~49--54, doi:\href{https://www.doi.org/10.26607/ijsl.v18i0.21}{10.26607/ijsl.v18i0.21}.

\bibitem[Bar{\'{a}} \em{et~al.}(2019)Bar{\'{a}}, Tapia, and Zamorano]{Bar2019}
Bar{\'{a}}, S.; Tapia, C.; Zamorano, J.
\newblock Absolute Radiometric Calibration of {TESS}-W and {SQM} Night Sky
  Brightness Sensors.
\newblock {\em Sensors} {\bf 2019}, {\em 19},~1336, doi:\href{https://www.doi.org/10.3390/s19061336}{10.3390/s19061336}.

\bibitem[Hart and Martinez(2006)]{Hart2006}
Hart, J.K.; Martinez, K.
\newblock Environmental sensor networks: A revolution in the earth system
  science?
\newblock {\em Earth Sci. Rev.} {\bf 2006}, {\em 78},~177--191.

\bibitem[Vairamani \em{et~al.}(2013)Vairamani, Mathivanan, Venkatesh, and
  Kumar]{Vairamani2013}
Vairamani, K.; Mathivanan, N.; Venkatesh, K.A.; Kumar, U.D.
\newblock Environmental parameter monitoring using wireless sensor network.
\newblock {\em Instrum. Exp. Tech.} {\bf 2013}, {\em
  56},~468--471.

\bibitem[Zhang \em{et~al.}(2004)Zhang, Sadler, Lyon, and Martonosi]{Zhang2004}
Zhang, P.; Sadler, C.M.; Lyon, S.A.; Martonosi, M.
\newblock Hardware design experiences in ZebraNet.
\newblock In Proceedings of the 2nd International Conference on Embedded
  Networked Sensor Systems, Baltimore, MD, USA, {\bf
 2004}; pp. 227--238.

\bibitem[Vera-Amaro \em{et~al.}(2019)Vera-Amaro, Angeles, and
  Luviano-Juarez]{Vera2019}
Vera-Amaro, R.; Angeles, M.E.R.; Luviano-Juarez, A.
\newblock Design and analysis of wireless sensor networks for animal tracking
  in large monitoring polar regions using phase-type distributions and single
  sensor model.
\newblock {\em IEEE Access} {\bf 2019}, {\em 7},~45911--45929.

\bibitem[Awadallah \em{et~al.}(2019)Awadallah, Moure, and
  Torres-Gonz{\'a}lez]{Awadallah2019}
Awadallah, S.; Moure, D.; Torres-Gonz{\'a}lez, P.
\newblock An internet of things (IoT) application on volcano monitoring.
\newblock {\em Sensors} {\bf 2019}, {\em 19},~4651, doi:\href{https://doi.org/10.3390/s19214651}{10.3390/s19214651}.


\bibitem[Alliance(2015)]{Lora2015}
Alliance, L.
\newblock \emph{A Technical Overview of LoRa and LoRaWAN};
\newblock White Paper; {\bf 2015}; {Volume 20}. \href{https://www.tuv.com/content-media-files/master-content/services/products/1555-tuv-rheinland-lora-alliance-certification/tuv-rheinland-lora-alliance-certification-overview-lora-and-lorawan-en.pdf}{Available online}, accessed on 15-06-2021.

\bibitem[Sornin \em{et~al.}(2015)Sornin, Luis, Eirich, Kramp, and
  Hersent]{Lorawan2015}
Sornin, N.; Luis, M.; Eirich, T.; Kramp, T.; Hersent, O.
\newblock \emph{LoRaWAN Specification};
\newblock {LoRa Alliance} {\bf 2015}. \href{https://osch.oss-cn-shanghai.aliyuncs.com/blogContentFileSnapshot/1556464676588.pdf}{Available online}, accessed on 15-06-2021.

\bibitem[Miao \em{et~al.}(2016)Miao, Zander, Sung, and Slimane]{Miao2016}
Miao, G.; Zander, J.; Sung, K.W.; Slimane, S.B.
\newblock \emph{Fundamentals of Mobile Data Networks}; Cambridge University Press: Cambridge, UK
, {2016}, doi:\href{https://doi.org/10.1017/CBO9781316534298}{10.1017/CBO9781316534298}.

\bibitem[Ams~AG(2018)]{Ams}
Ams~AG
\newblock {\em TSL237 High-Sensitivity Light-to-Frequency Converter} {March 2018}; {v1-01}. \href{https://ams.com/documents/20143/36005/TSL237_DS000156_3-00.pdf/4aa35672-5c5e-3bb7-4d6b-92f4c76a3531}{Available online}, accessed on 28-06-2021.

\bibitem[Pravettoni \em{et~al.}(2016)Pravettoni, Strepparava, Cereghetti,
  Klett, Andretta, and Steiger]{Pravettoni2016}
Pravettoni, M.; Strepparava, D.; Cereghetti, N.; Klett, S.; Andretta, M.;
  Steiger, M.
\newblock Indoor calibration of Sky Quality Meters: Linearity, spectral
  responsivity and uncertainty analysis.
\newblock {\em J. Quant. Spectrosc. Radiat. Transf.}
  {\bf 2016}, {\em 181},~74--86, doi:\href{https://www.doi.org/10.1016/j.jqsrt.2016.03.015}{10.1016/j.jqsrt.2016.03.015}.

\bibitem[{Bessell}(1990)]{Bessell1990}
{Bessell}, M.S.
\newblock {UBVRI passbands.}
\newblock {\em Publ. Astron. Soc. Pac.} {\bf 1990}, {\em 102},~1181--1199, doi:\href{https://www.doi.org/10.1086/132749}{10.1086/132749}.

\bibitem[{Noll} \em{et~al.}(2012){Noll}, {Kausch}, {Barden}, {Jones},
  {Szyszka}, {Kimeswenger}, and {Vinther}]{Noll2012}
{Noll}, S.; {Kausch}, W.; {Barden}, M.; {Jones}, A.M.; {Szyszka}, C.;
  {Kimeswenger}, S.; {Vinther}, J.
\newblock {An atmospheric radiation model for Cerro Paranal. I. The optical
  spectral range}.
\newblock {\em Astron. Astrophys.} {\bf 2012}, {\em 543},~A92, doi:\href{https://doi.org/10.1051/0004-6361/201219040}{10.1051/0004-6361/201219040}.

\bibitem[Cinzano(2005)]{Cinzano2005}
Cinzano, P.
\newblock \emph{Night Sky Photometry with Sky Quality Metter}. {\bf 2005}. \href{https://www.researchgate.net/publication/228399779_Night_Sky_Photometry_with_Sky_Quality_Meter}{Available online}, accessed on 28-06-2021.

\bibitem[Zamorano \em{et~al.}(2017)Zamorano, Tapia, Garc\'{i}a, Gonz\'{a}lez,
  de~Miguel, Bar\'{a}, S.~Pascual, Garc\'{i}a, and
  consortium]{PosterZamorano2017}
\textls[-15]{Zamorano, J.; Tapia, C.; Garc\'{i}a, C.; Gonz\'{a}lez, R.; de~Miguel, A.S.;
  Bar\'{a}, S.; Pascual, J.G.S.; Garc\'{i}a, L.; consortium, T.}
\newblock Poster: Calibration of {TESS}, the {STARS4ALL} night sky brightness
  photometer.
\newblock  In \emph{Light Pollution: Theory, Modelling and Measurements}; {\bf 2017}. \href{https://guaix.ucm.es/wp-content/uploads/2020/02/2017_06_LPTMM_poster_TESS_calibration.pdf}{Available online}, accessed on 28-06-2021.


\end{thebibliography}


\end{document}